\documentclass[journal]{IEEEtran}

\usepackage{xcolor,soul,framed} 

\colorlet{shadecolor}{yellow}
\usepackage{graphicx}
\graphicspath{{../pdf/}{../jpeg/}}
\DeclareGraphicsExtensions{.pdf,.jpeg,.png}

\usepackage[cmex10]{amsmath}
\usepackage{array}
\usepackage{mdwmath}
\usepackage{mdwtab}
\usepackage{eqparbox}
\usepackage{url}
\usepackage[numbers,sort&compress]{natbib}
\usepackage{comment}

\begin{document}
\bstctlcite{IEEEexample:BSTcontrol}
    \title{Optoelectronic recurrent neural network using optical--electrical--optical converters with RC delay}
  \author{Masaya~Arahata,~\IEEEmembership{}
      Shota~Kita,~\IEEEmembership{}
      Kazuo~Aoyama,~\IEEEmembership{}
      Akihiko Shinya,~\IEEEmembership{}
      Hiroshi Sawada,~\IEEEmembership{}
      and~Masaya~Notomi,~\IEEEmembership{}
  }  
\maketitle

\begin{abstract}
Optical neural network (ONN) has been attracting intense attention owing to their low latency and low-power consumption. Among the ONNs, optical recurrent neural network (RNN) enables low-power and high-speed time-series data processing using a compact loop structure. The loop losses need to be efficiently compensated so that the time-series information is maintained in the RNN operation. For this purpose, we focus on the optoelectronic RNN (OE-RNN) with optical--electrical--optical (OEO) converters to compensate for the loop losses. However, the effect of resistive--capacitive (RC) delay of OEO converters on the RNN performance is unclear. Here, we study in simulation an OE-RNN equipped with OEO converters with RC delay. We confirm that our modeled OE-RNN achieves the high training accuracy of time-series data classification even when RC delay is comparably large to the time interval of time-series data. Our analyses reveal that the accumulation of time-series data by RC delay does not degrade the RNN performance but rather can compensate for the degraded RNN performance due to loop losses. From the theoretical analysis referring to the gradient explosion and vanishing problems, we find the region related to loss and RC delay where the high training accuracy can be achieved. In simulation, we confirm this compensation effect in the large OE-RNN circuit up to 32$\times$32 scale. Our proposed scheme opens a new way of time-series data processing by utilizing RC delay for the optical computing and optical communication.   
\end{abstract}


\IEEEpeerreviewmaketitle


\section{Introduction} \label{introduction}

\IEEEPARstart{M}{achine} learning (ML) is a powerful tool for the computation and classification of big data such as speech recognition and automatic driving, which became essential techniques in the modern society \cite{LeCun2015,Assael2022}. The ML based on a neural network (NN) can efficiently process the large data by using a graphics processing unit (GPU). However, the energy consumption of resources required for processing the huge amount of data is quite large. The power consumption of NN inference or learning increases mainly due to the multiply-accumulate (MAC) operations needed to perform matrix-vector multiplication \cite{Sze2017,Patterson2021}. 

Recently, an optical neural network (ONN) \cite{Shen2017,Tait2017,Feldmann2021} has been attracting intense attention owing to its low latency and low-power consumption based on analog computing. The ONN has a strong point of multiplexing utilizing a degree of freedom of light such as spatial \cite{Ikeda2024,Onodera2024}, wavelength \cite{Feldmann2021,Xu2021,Xu2023}, and temporal domain \cite{Bueno2018,Dong2023}. 

Among the ONNs, an optical recurrent neural network (RNN) with an internal feedback loop structure has strong advantages that (1) it can work with the high-speed and low-power consumption in sequential data processing tasks such as speech recognition, natural language processing, and financial forecasting and (2) small-scale circuits with the RNN structure can process the high-dimensional data by expanding it into the time domain. However, optical RNN has a serious problem that the RNN performance is degraded by the light attenuation due to internal loop losses. So far, all-optical RNNs were demonstrated for applications such as a gated recurrent unit (GRU) or a solver of integro-differential equation \cite{Alexandris2020,Feng2020,Chen2024,Liu2024}. In these demonstrations, a semiconductor optical amplifier (SOA) or an erbium doped fiber amplifier (EDFA) were used outside a chip to compensate for the loop losses in the RNN circuit. Such amplifiers are still difficult to be implemented on a Si photonic chip due to the heat and noise issues.

To deal with this problem, we focus on an optoelectronic RNN (OE-RNN) which can compensate for the loop losses by using optical-electrical-optical (OEO) converters, where the electrical signal converted from input optical signal drives a modulator to control an external optical signal. The OEO converters used in the OE-RNN have several advantages that they (1) have much less heat (scalable) and intrinsically no amplified spontaneous emission (ASE) noise by using an amplifier-free \cite{nozaki2016} and bias-free operation \cite{nozaki2018}, (2) have the gain for the input signal suffering from the loop losses, and (3) act as an activation function owing to their nonlinearity \cite{Bandyopadhyay2022}. Recent advances in integration technology of Si photonics lead to the feasibility of the on-chip OEO converter \cite{Bandyopadhyay2022}, which is more compact and has the lower latency and lower power consumption than a bulky off-chip OEO converter \cite{Tait2017}. Since the OE-RNN can be compatibly connected with the nanophotonics, the OEO converters can be miniaturized with the ultra-low capacitance for the ultra-low power consumption operation \cite{nozaki2019}. Even with the small capacitance, the large load resistance is desirable for the improvement of effective OEO gain. In addition, the OE-RNN with a larger resistive-capacitive (RC) constant is generally easier to fabricate in terms of the device integration technologies \cite{Siew2021}. Therefore, the analysis of the effect of RC delay on ONNs is important for designing actual circuits for both ultimate and practical purposes. However, there has not been quantitative studies of the effect of RC delay for ONNs as far as we know.

In this paper, we investigate the OE-RNN equipped with OEO converters with the RC delay and show that the RC delay is not an obstacle to the OE-RNN operation. We confirm that our modeled OE-RNN achieves the high training accuracy of time-series data classification even when RC delay is comparably large to the time interval of time-series data. Surprisingly, our analyses reveal that the accumulation of time-series data by RC delay does not degrade the RNN performance but rather can compensate for the degraded RNN performance due to loop losses. From the theoretical analysis referring to the gradient explosion and vanishing problems, we find the region related to loss and RC delay where the high training accuracy can be achieved. In simulation, at first, we perform the image classification of MNIST task as sequential data expanded to the time domain using a 4$\times$4 OE-RNN circuit with OEO converters. We assume the OEO converters to be photodetectors (PD) and EOM, and devise a general linear model based on the impulse response that takes into account their RC delay. Our simulation results show that the training accuracy of the sequential MNIST task by our modeled OE-RNN increases with the RC delay when loop losses exist, which is in good agreement with our analysis. Next, we perform the simulation on the larger-scale OE-RNN circuit up to $32\times32$ to confirm the compensation effect of RC delay of OEO converters. The results clearly show that the RC delay of OEO converters can compensate for the degraded accuracy of OE-RNN due to the gradient explosion or vanishing problems.

The rest of the paper is constructed as follows. In Section \ref{analysis}, we present a general analysis approach of the effect of RC delay of OEO converters on the OE-RNN performance. Section \ref{simulation} describes the simulation results of OE-RNN equipped with OEO converters with up to 32$\times$32 scale. In Section \ref{discussion}, we discuss the experimental feasibility of pulse response of OEO converters. Section \ref{conclusion} summarizes the novelty of this work and presents the conclusions.

\section{Analysis of the effect of RC delay of OEO converters on the OE-RNN performance} \label{analysis}

\begin{figure*}
  \begin{center}
  \includegraphics[width=6.5in]{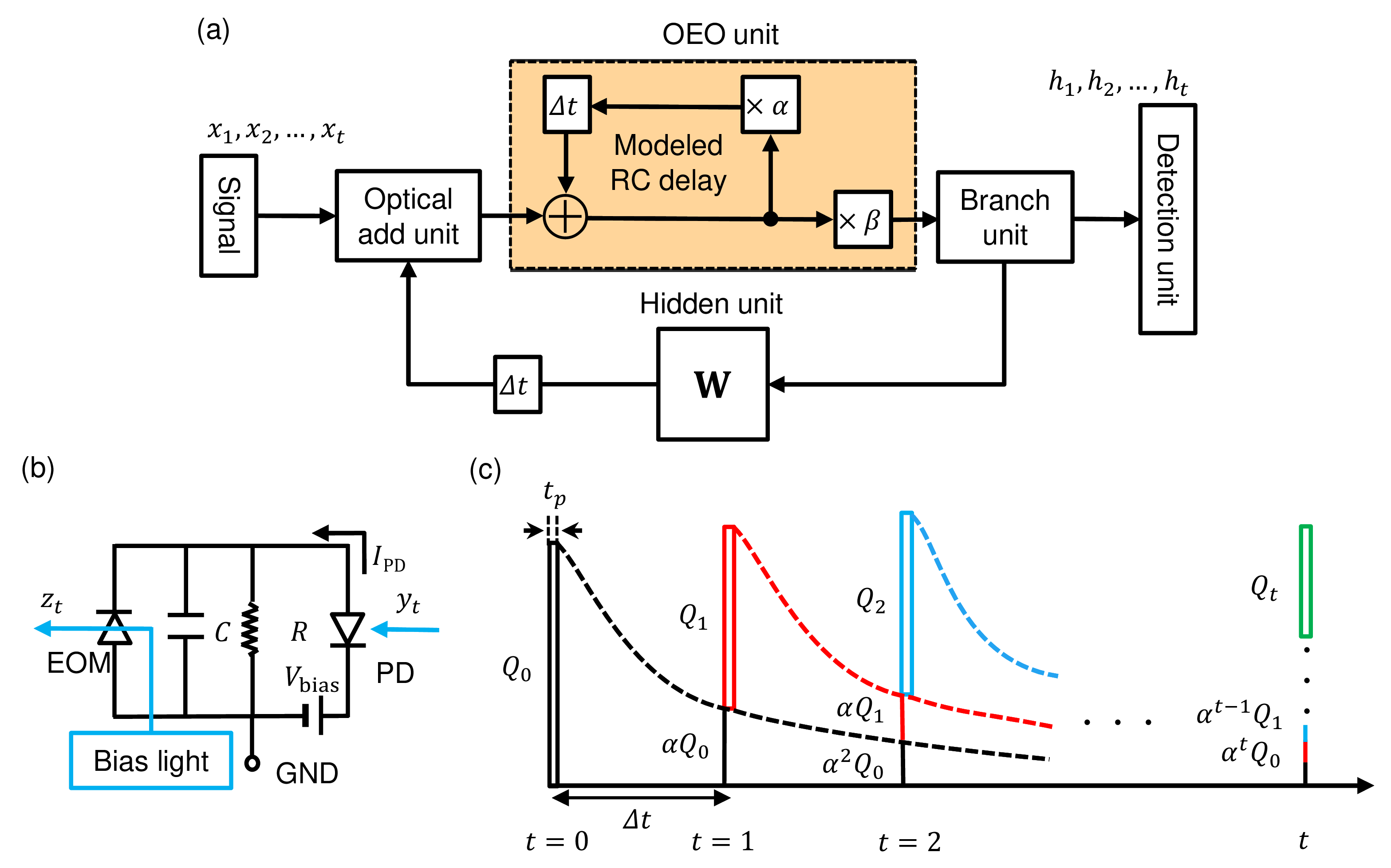}\\
  \caption{(a) Schematic diagram of Elman-type OE-RNN with the RC delay of OEO converter. (b) Model of OEO converter with the RC delay, respectively. W: hidden unit, $I_{\rm{PD}}$: photocurrent, $V_{\rm{bias}}$: bias voltage, EOM: electro-optic modulator. (c) Scheme of the accumulation of charge $Q$ of EOM capacitance by RC delay in the case of impulse response.}\label{fig_OEO_model}
  \end{center}
\end{figure*}

In this section, we describe the setup of OE-RNN circuit with the OEO converter and the analysis of the effect of RC delay of OEO-converter on the OE-RNN performance. Fig. \ref{fig_OEO_model}(a) shows the schematic diagram of Elman-type OE-RNN equipped with an OEO converter with the RC delay. The circuit consists of a signal input unit, an optical add unit, a branch unit, a detection unit, Clements circuit \cite{Clements2016} as a hidden unit of unitary matrix $W$ in the loop, and a modeled OEO unit, where an output unit and activation function are not included. The learning parameters are the phases of phase shifters in the optical add unit, branch unit, and hidden unit. The complex optical input signal $x_{t}$ at the discrete time $t$ ($\geq1$) is injected into the OE-RNN circuit. The output signal of optical add unit $y_{t}$ enters the OEO converter and then the complex optical output signal $h_{t}$ is branched to the output of network and the input of hidden unit $W$ in the optical loop. The output signal of hidden unit is coherently or incoherently added to the next input signal whose wavelength is same or different as the output signal of hidden unit, respectively. We mathematically model the OEO converter where the current complex optical input signal $y_{t}$ to the OEO converter is amplified by a factor of the loop gain $\beta$ and the previous complex output signal $z_{t-1}$ by the exponential RC delayed feedback $\alpha$ is added to it by the accumulation of time-series information due to RC delay expressed as Eq. \ref{OEO_model}. The circuit of OEO converter is depicted in Fig. \ref{fig_OEO_model}(b). The input light signal $y_{t}$ to the OEO converter is detected by a PD and then the photocurrent $I_{\rm{PD}}$ is converted into the voltage by a load resistance $R$. The load voltage drives an EOM to modulate an external bias light with the same or different wavelength as the input light for coherent or incoherent operation, respectively, resulting in the output signal $z_{t}$. Since in practice the EOM and PD has the finite capacitance, the optical and electrical signals are affected by the RC delay. Here, we assume that the input signal is a short pulse to the RC time constant of OEO converters (pulse width $t_{\rm {p}} \ll$ RC time constant $\tau_{RC}$) and the loop time is equal to the time interval of data $\Delta t$. When the RC time constant is larger than the time interval of data ($\Delta t \ll \tau_{RC}$), the decayed previous signals by RC delay remain after the next loop time as shown in Fig. \ref{fig_OEO_model}(c). Since our OEO converter is a linear time-invariant system consisting of an RC circuit, the charge of capacitance at time $t$ is superposed by $Q_{t}+\alpha Q_{t-1}+\cdots+\alpha^{t-1}Q_{1}$ in discrete time domain, where the charge of capacitance by the input signal is $Q_{t}=y_{t}t_{p}$ and $\alpha \equiv \exp(-\Delta{t}/\tau_{RC})$ is the typical exponential RC delayed feedback with time constant which $\tau_{RC}\rightarrow0$, $\alpha\rightarrow0$ and $\tau_{RC}\rightarrow\infty$, $\alpha\rightarrow1$. Here we assume a Mach-Zehnder modulator (MZM) as an EOM. In the case of direct detection of input signal by a PD, the signal power is converted into the charge to the capacitance of OEO converter and its phase information is lost. With the assumption that the complex number information of input signal is kept by using a coherent detection, the relation of optical intensity between the input and output of OEO converter is expressed as
\begin{equation}\label{MZM_output_coherent}
\begin{split}
z_{t} &= E_{\rm bias}\sin{\left(\frac{\pi\beta_{0}t_{\rm {p}}}{2V_{\pi}C}\sum_{d=0}^{t-1}{\alpha^{d}y_{t-d}}\right)} \\
&\approx \frac{E_{\rm bias}\pi\beta_{0}t_{\rm {p}}}{2V_{\pi}C}\sum_{d=0}^{t-1}{\alpha^{d}y_{t-d}},
\end{split}
\end{equation}
where $\beta_{0}$ is the loop loss for the optical amplitude, $E_{\rm bias}$ is the optical signal of bias light, and $V_{\pi}$ is the $\pi$-shift voltage, and for the simplicity, we assume that the MZM is operated in the linear region with the small input signal. Totally, the Eq. \ref{MZM_output_coherent} is rewritten simply
\begin{equation}\label{OEO_model}
z_{t} = \beta y_{t} + \alpha z_{t-1},
\end{equation}
where $\beta=G\beta_{0}$ and $G=E_{\rm bias}\pi t_{\rm {p}}/2V_{\pi}C$ is the loop gain of OE-RNN system and gain of OEO converter, respectively. The OEO converter acts as a gain or loss for the optical input signal when $\beta > 1$ or when $\beta < 1$, respectively. Eq. \ref{OEO_model} shows that the output signal of OEO converter is affected by the input signal at the current time and the output signal at the previous loop time by the accumulation of time-series information due to the RC delay.

Next, to investigate the behavior of OE-RNN circuit with RC delay, we formulate the relation between the input and output signal in OE-RNN circuit. From Fig. \ref{fig_OEO_model}(a), (b), and Eq. \ref{OEO_model}, the relation between the input and output signals of OE-RNN circuit at loop time $t$ is expressed as
\begin{equation}\label{RNN_output}
h_{t} = \beta(x_{t} + Wh_{t-1})+\alpha h_{t-1} = \beta x_{t} + (\beta W + \alpha I)h_{t-1},
\end{equation}
where $W$ and $I$ denote the matrix of the hidden unit and an identity matrix, respectively. Thus, the output signal is expressed with the input signal at each time as 
\begin{equation}\label{RNN_output_input}
h_{t} = \beta\sum\limits_{k=1}^t S^{t-k} x_{k},
\end{equation}
where $S\equiv\beta W + \alpha I$. Note that Eq. \ref{RNN_output_input} shows that the output of OE-RNN is influenced by each time-series input history with the different number of multiplication of matrix $S$. When loop gain $\beta=1$ and RC delayed feedback $\alpha=0$, OE-RNN can process the time-series data in the same manner as the conventional RNN.

Here, we analyze the effect of the RC delay of OEO converter on the OE-RNN performance referring to the gradient explosion and vanishing problems \cite{Arjovsky2016,Wisdom2016,Jing2016}. Their recently proposed RNN with a unitary matrix as a hidden unit can keep the gradient stable during training since all its eigenvalues have absolute values of unity. Here, we show that in the case of OE-RNN, the gradient explosion or vanishing caused by the deterioration of unitarity of matrix $S$ due to the loop gain or loss can be compensated by the RC delay of OEO converter. When training the neural network to minimize a loss function $L$ that depends on parameters of $W$, the gradient vanishing and explosion problems refer to the decay or growth of $\frac{\partial L}{\partial h_{t}}$ as the number of layers $t$ increases. By using the chain rule,
\begin{equation}\label{cost_gradient}
    \begin{aligned}
        \frac{\partial L}{\partial h_{t}} &= \frac{\partial L}{\partial h_{\rm {T}}}\frac{\partial h_{\rm {T}}}{\partial h_{t}} \\
        &= \frac{\partial L}{\partial h_{\rm {T}}}\prod_{k=t}^{\rm {T}-1} \frac{\partial h_{k+1}}{\partial h_{k}} \\
        &= \frac{\partial L}{\partial h_{\rm {T}}}\prod_{k=t}^{\rm {T}-1} S,
    \end{aligned}
\end{equation}
where $T$ is a sequential length of input data. For large lengths $T$, the term $\prod S$ plays a significant role. Then if $S$ has eigenvalues $|\lambda_{\rm {s}}| \ll 1$, they will cause gradient vanishing $|\frac{\partial L}{\partial h_{t}}| \rightarrow 0$, while if $S$ has eigenvalues $|\lambda_{\rm {s}}| \gg 1$, they can cause gradient explosion $|\frac{\partial L}{\partial h_{t}}| \rightarrow \infty$. Both the situations prevent the RNN from working efficiently. Here, the eigenvalues $\lambda_{\rm {s}}$ can be calculated as
\begin{equation}\label{eigenvalues}
    \begin{split}
    S\vec{v}_{h}&=(\beta W+\alpha I)\vec{v}_{h} \\
            &=(\beta \lambda_{h}+\alpha)\vec{v}_{h} \\
            &=\lambda_{\rm {s}}\vec{v}_{h},
    \end{split}
\end{equation}
where $\vec{v}_{h}$ $(h=1,2,\cdots,N)$ are the eigenvectors of $N\times N$ unitary matrix $W$ and $S$ and $\lambda_{h}=e^{i\theta}$ $(0\leq\theta\leq2\pi)$ are the eigenvalues of $W$ for each eigenvector whose absolute values are exactly 1. From $0\leq\alpha\leq1$ and $\beta\geq0$, the range of absolute value of $\lambda_{\rm {s}}$ is $|\alpha -\beta|\le|\lambda_{\rm {s}}|\le \alpha + \beta$, so the necessary condition of $\alpha$ and $\beta$ where $|\lambda_{\rm {s}}|=1$ is expressed as
\begin{equation}\label{alpha_beta_condition}
-\alpha + 1 \le \beta \le \alpha + 1.
\end{equation}
When the above inequality is satisfied, the gradient explosion and vanishing can be relaxed. These analyses show that even when the unitarity of weight matrix is degraded by the loss or gain of RNN circuit, the unitary condition can be recovered by appropriately adding the diagonal term of RC delay of OEO converter. Therefore, if the OEO converter has the gain or loss to cause the gradient explosion or vanishing problems, controlling the RC delay of OEO converter enables to prevent them.

\section{Simulation results of OE-RNN equipped with OEO converters} \label{simulation}

\subsection{$4\times4$ OE-RNN}

\begin{figure*}
  \begin{center}
  \includegraphics[width=6.7in]{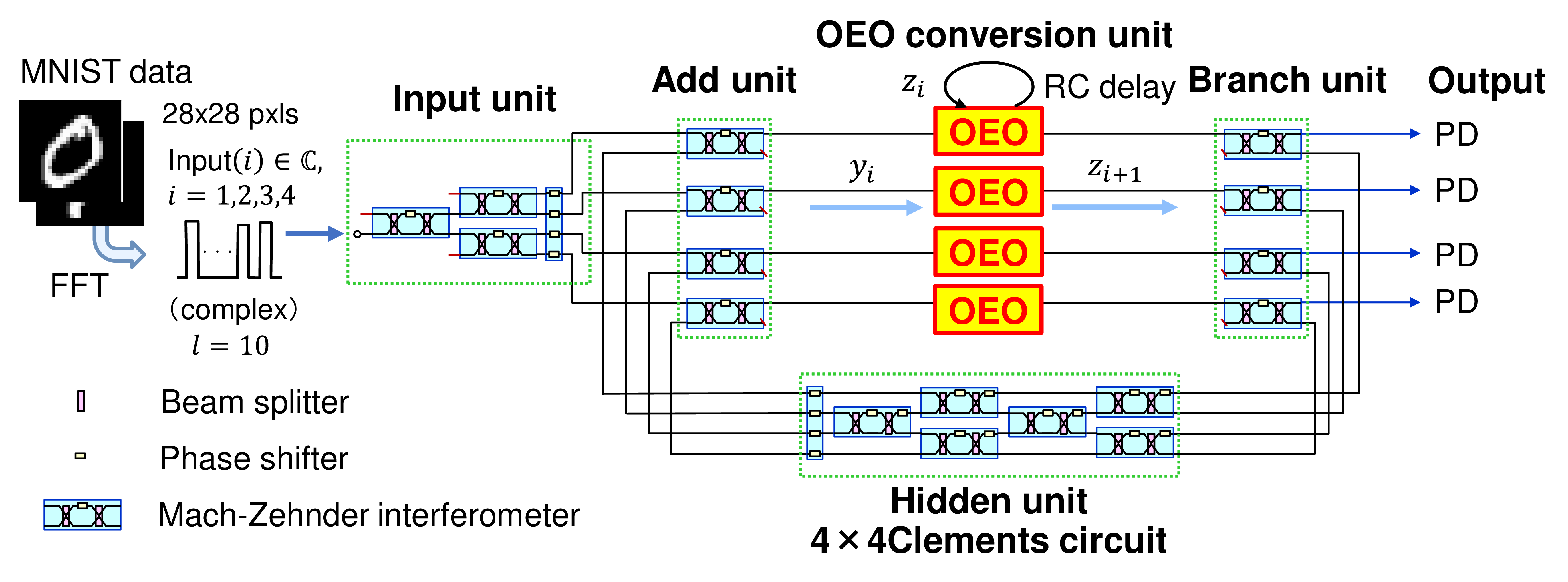}\\
 \caption{Schematic diagram of Elman-type $4 \times 4$ OE-RNN circuit. The input, hidden, and branch units are composed of MZIs with 50/50 BSs and PSs. After FFT, time-series input signals for each MNIST image are generated from its selected FFT signals. The relation between the input and output of OEO unit is expressed by Eq. \ref{OEO_model} at time $t$ and $t-1$ with the loop gain $\beta$ and RC delayed feedback $\alpha$. The output signal of branch unit is detected by PDs. While the RNN learning, the phase of PS is tuned for minimizing the loss function.}\label{RNN_diagram_4_4}
  \end{center}
\end{figure*}

To demonstrate the analysis of the effect of RC delay of OEO converters on the OE-RNN performance, we performed the simulation of OE-RNN circuit with the RC delay and loop losses. Fig. \ref{RNN_diagram_4_4} shows a schematic diagram of Elman-type $4 \times 4$ OE-RNN circuit with OEO converters. To perform the sequential MNIST task with four-digits of \{0,1,2,3\} using a $4 \times 4$ OE-RNN circuit, we preprocess input images of MNIST data into those with appropriate size. Performing Fast Fourier Transform (FFT) of four-digits input images with $28\times28$ pixels, we select top-10 components in descending order of power from the compressed 393 signals (392 and DC signals) as the time-series inputs with a sequential length of $l=10$. Note that OE-RNN can process the larger-scale data by expanding it into the time domain than one that the conventional feedforward NN can handle. Here, input, hidden, and branch units are composed of Mach-Zehnder interferometers (MZI) with 50/50 beam splitters (BS) and phase shifters (PS). In the input unit, the one dimensional input data signal at the time is converted into an $N$-dimensional vector each element of which is multiplied by the corresponding weight. The hidden unit is $4 \times 4$ Clements circuit that is capable of expressing a $4 \times 4$ unitary matrix. In the branch unit, the power ratio of light to the loop and output is fixed to 0.96:0.04 to reduce the additional loop losses. The OEO unit has the loop gain $\beta$ and RC delayed feedback $\alpha$. The output signal of hidden unit is added to the next input signal in the add unit. The output signal of branch unit is detected by PDs. The detected output is used for the loss function calculation, which is defined as the cross entropy after passing through a softmax function. While the RNN learning, the phase of PS is tuned for minimizing the loss function.

\begin{figure*}
  \begin{center}
  \includegraphics[width=7.2in]{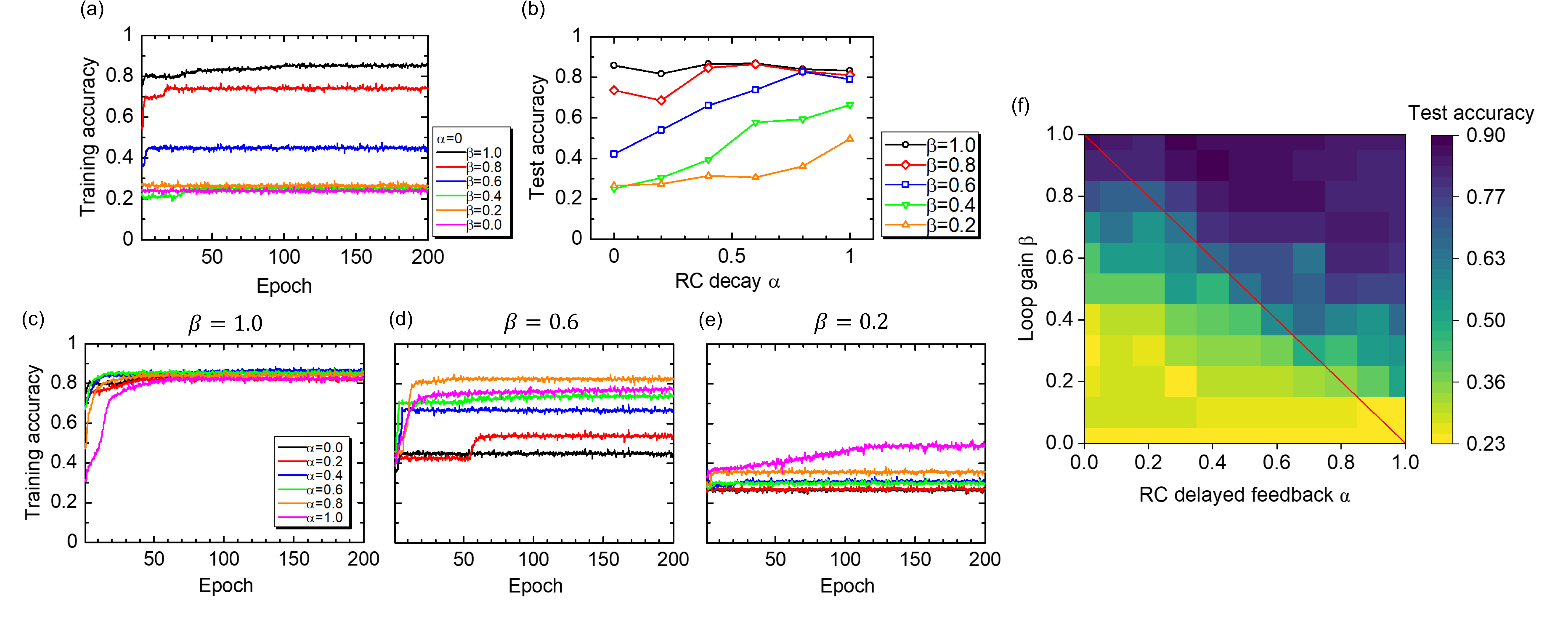}\\
 \caption{Simulation results of Elman-type $4 \times 4$ OE-RNN circuit. (a) Loop gain $\beta$ dependence of training accuracy with the RC delayed feedback $\alpha=0$. (b) RC delayed feedback dependence of test accuracy with each loop gain. RC delayed feedback dependence of training accuracy with the loop gain (c) $\beta=1.0$, (d) $\beta=0.6$, and (e) $\beta=0.2$, respectively. (f) RC delayed feedback $\alpha$ and loop gain $\beta$ dependence of test accuracy, where the step size of $\alpha$ and $\beta$ is 0.1 and the red line corresponds to $\beta=-\alpha+1$ in Eq. \ref{alpha_beta_condition}.}\label{4_4_results}
  \end{center}
\end{figure*}

Fig. \ref{4_4_results} shows the learning results of the OE-RNN in Fig. \ref{RNN_diagram_4_4}. Fig. \ref{4_4_results}(a) shows the loop gain $\beta$ dependence of training accuracy of $4 \times 4$ OE-RNN circuit with no RC delay ($\alpha=0$). With the loop gain $\beta=1$, the high training accuracy of about 80\% was achieved even under the condition that the pixel size of 784 of input MNIST data were compressed to 10. We can see that as the loop gain decreased ($\beta < 1$), the training accuracy degraded. This degradation is caused by the loss of time-series information of input data due to the loop losses corresponding to the gradient vanishing. Fig. \ref{4_4_results}(b) shows the RC delayed feedback $\alpha$ dependence of test accuracy at each loop gain and Fig. \ref{4_4_results}(c-e) show the training accuracy at (c) $\beta=1.0$, (d) $\beta=0.6$, and (e) $\beta=0.2$ of $4 \times 4$ OE-RNN circuit, respectively. In Fig. \ref{4_4_results} (c), we can see that almost the same training accuracy of 80\% is achieved as the RC delayed feedback $\alpha$ increases. This result shows that the our modeled OE-RNN achieves the high training accuracy without a loss ($\beta = 1$) even when RC delay is comparably large to the time interval of data ($\tau_{RC}\sim\Delta t$ corresponding to $\alpha\approx0.4$). On the other hand, in Fig. \ref{4_4_results} (d) and (e), we can see the improvement of the training accuracy by increasing RC delayed feedback $\alpha$ even when the large loop losses exist ($\beta < 1$) also shown in Fig. \ref{4_4_results} (b). This is caused by the compensation of degraded accuracy due to loop losses by the accumulation of time-series information by RC delay as predicted from the analysis in Section \ref{analysis}. These results clearly show that the RC delay of OEO converters does not degrade the RNN performance but rather can compensate for the degrade RNN performance due to loop losses.

To quantitatively analyze the effect of the RC delay and loop gain on the RNN performance, we investigate the RC delayed feedback $\alpha$ and loop gain $\beta$ dependence on the test accuracy of MNIST task as shown in Fig. \ref{4_4_results}(f). The step size of $\alpha$ and $\beta$ is 0.1. The result clearly shows that the accuracy increases with the large $\alpha$ when $\beta$ is small. Furthermore, we can see that in the area above the red line corresponding to the left hand side of Eq. \ref{alpha_beta_condition}, the test accuracy is higher than that in the outside of area. These results clearly show that the RC delay can prevent the gradient vanishing to compensate for the degrade RNN performance in particular when the loop losses are large.

\subsection{Larger-scale OE-RNN}

\begin{figure*}
  \begin{center}
  \includegraphics[width=6.0in]{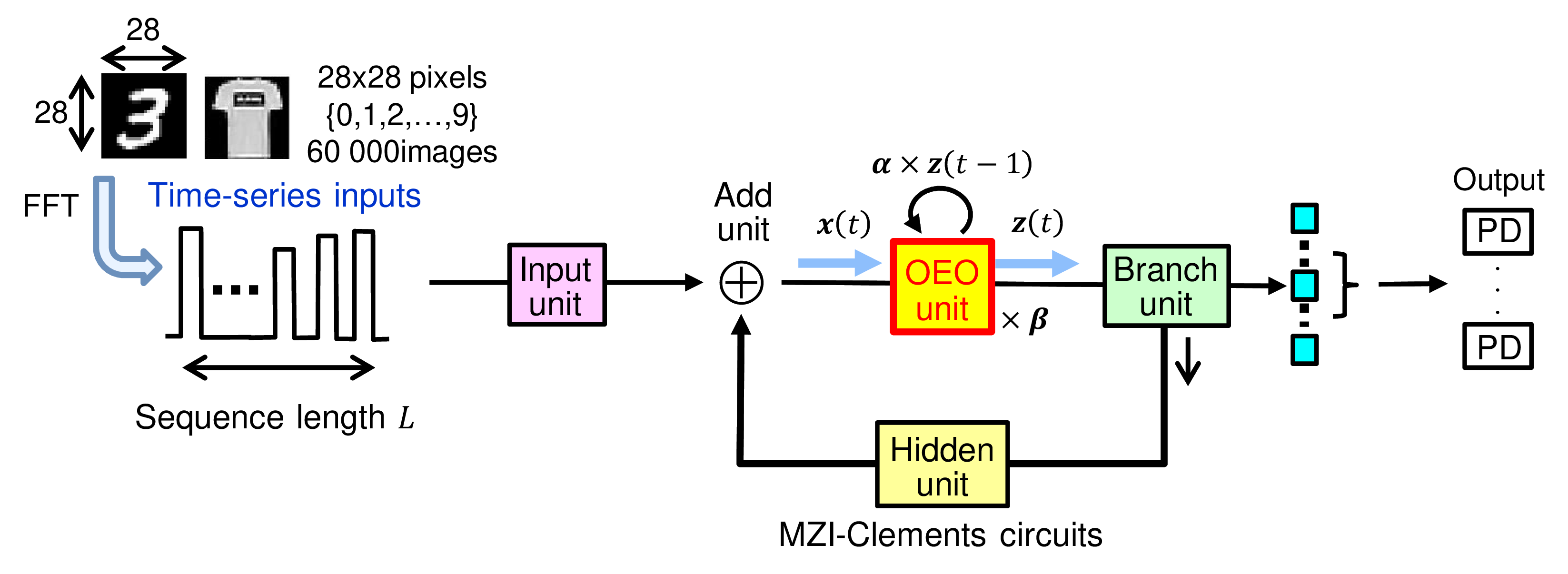}\\
 \caption{Schematic diagram of Elman-type $N \times N$ OE-RNN circuit. The one dimensional time-series input data is distributed into the N port by using the branch structure in the input unit.  The OEO unit has the loop gain $\beta$ and RC delayed feedback $\alpha$. The output of hidden unit composed of $N\times N$ Clements circuit is coherently added to the next input data. The selected output signal of branch unit around the center for the classification digit is detected by PDs.}\label{large_RNN_diagram}
 \end{center}
\end{figure*}

To confirm the effect of RC delay of OEO converters on the performance of larger-scale OE-RNN, we simulate the larger-scale OE-RNN circuit. Fig. \ref{large_RNN_diagram} shows a schematic diagram of Elman-type $N\times N$ ($N\geq8$) OE-RNN circuit with OEO converters. The hidden unit is $N\times N$ Clements circuit that is capable of expressing an $N\times N$ unitary matrix. For the input with the larger sequential length of $l$, the one dimensional data signal is distributed into the $2^{m}=N$ dimensional using $2^{m}-1$ MZI with the branch structure in the input unit. The OEO unit has the loop gain $\beta$ and RC delayed feedback $\alpha$. In the branch unit, the power ratio of light to the loop and output is fixed to 0.96:0.04 to reduce the additional loop losses. The output signal of hidden unit is coherently added to the next input signal. The selected 10 output signals of branch unit around the center are detected by PDs. 

\begin{figure*}[t]
  \begin{center}
  \includegraphics[width=7.0in]{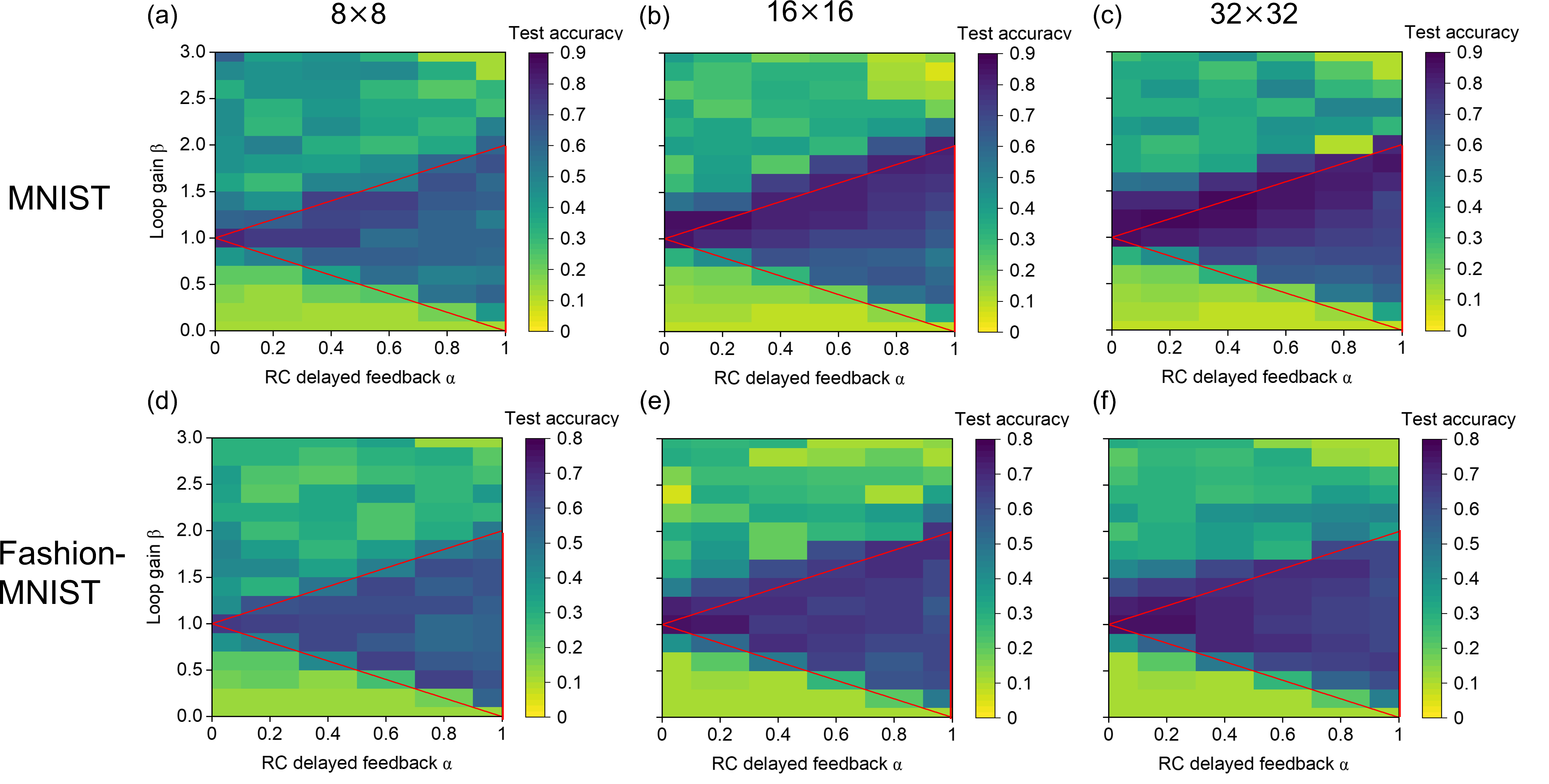}\\
 \caption{RC delayed feedback $\alpha$ and loop gain $\beta$ dependence of test accuracy of sequential (a-c) MNIST and (d-f) Fashion-MNIST task with $l=20$ of (a,d) $8 \times 8$, (b,e) $16\times16$, and (c,f) $32\times32$ OE-RNN circuit, respectively. The area surrounded by red lines corresponds to Eq. \ref{alpha_beta_condition}.}\label{Larger_scale_MNIST}
  \end{center}
\end{figure*}

Fig. \ref{Larger_scale_MNIST} shows the RC delayed feedback $\alpha$ and loop gain $\beta$ dependence of test accuracy of sequential (a-c) MNIST and (d-f) Fashion-MNIST task with $l=20$ of (a,d) $8 \times 8$ for 8-digit, (b,e) $16\times16$ for 10-digit, and (c,f) $32\times32$ for 10-digit OE-RNN circuit, respectively. The step sizes of $\alpha$ and $\beta$ are 0.2 in all data. We can see that the test accuracy increases in the area surrounded by red lines corresponding to Eq. \ref{alpha_beta_condition}. On the contrary, the test accuracy is lower outside the area. This reflects the fact that adjusting the relation between the RC delayed feedback $\alpha$ and loop gain $\beta$ enables to avoid the gradient explosion or gradient vanishing problem. We can see clearly this effect in the simulation results with the larger circuit size for not only MNIST but also more difficult Fashion-MNIST task, resulting that our analysis is valid for the hard time-series information task. The simulation results are in good agreement with the theoretical analysis. These results clearly show that the RC delay of OEO converter can compensate for the degraded training accuracy of OE-RNN due to the excessive gain or large loop losses to prevent the gradient explosion or vanishing problem.

Although we assumed the OE-RNN with a unitary matrix as the hidden unit, our proposed method to evaluate the effect of RC delay of OEO converters can also be applied to different types of matrices. For example, when we use a real-valued ONN circuit such as using an incoherent microring resonator weight bank \cite{Tait2014}, the analysis in Section \ref{analysis} can be applied by replacing the complex amplitude of signal lights in OE-RNN and the unitary matrix with the intensity of them and the orthonormal matrix whose eigenvalues are $\pm1$, respectively. The circuit matrix is defined as $S'\equiv\beta W' + \alpha I$ where $W'$ is the orthonormal matrix in Eq. \ref{RNN_output_input}. Totally, the necessary condition of $\alpha$ and $\beta$ for preventing the gradient explosion and vanishing where absolute eigenvalues of $S'$ are 1 ($|\lambda_{\rm {S'}}|=1$) is equal to Eq. \ref{alpha_beta_condition}. For a non-unitary complex matrix as the hidden unit, the necessary condition of $\alpha$ and $\beta$ where the absolute values of eigenvalues of circuit matrix are equal to 1 is different to Eq. \ref{alpha_beta_condition}, which depends on the eigenvalues of the complex matrix. From the similar analysis, we can obtain the inequality of loop gain $\beta$ and RC delayed feedback $\alpha$ where the gradient explosion and vanishing are relaxed. Therefore, the RC delay of OEO converter could improve the performance of OE-RNN with not only Clements circuit but also other types of hidden unit.

\section{Discussion} \label{discussion}

\subsection{Feasibility of pulse response of OEO converter}

In this section, to verify the experimental feasibility of OEO model expressed as Eq. \ref{OEO_model}, we discuss the pulse response of OEO converter based on the device parameters. First, we investigate the impulse response of OEO converter. With the assumption of impulse response, since the charge to an EOM by input signal is very rapid, the voltage applied to EOM $V_{\rm{EOM}}$ is reduced to $I_{x}t_{p}/C$, where $I_{x}$ is the current to EOM corresponding to a peak power of the input signal of OEO converter, compared to the case of finite-pulse response as $V_{\rm{EOM}}=RI_{x}$. Therefore, the larger peak power of input signal is required for the $\pi$ phase shift of EOM expressed as $I_{x}=CV_{\pi}/t_{p}$. From the condition of impulse response as $t_{p} \ll \tau_{RC}$, $I_{x} \gg V_{\pi}/R$. Table \ref{OEOtable} shows the comparison of device parameters of four kinds of OEO converters in the case of impulse response. Here, we set the values of load resistance as 5 k$\Omega$ and assume that the O-E conversion is performed by a Ge PD with a typical value of capacitance of 5 fF. We can see in Table \ref{OEOtable} that for OEO converters with the smaller capacitance (smaller RC time constant), the short pulse width is needed for the validation of impulse response $t_{p} \ll \tau_{RC}$ while the small peak power of input light is sufficient for the operation. For all the OEO converters, it is feasible for the current technology to use the pulse laser with the pulse width of from sub-ps to sub-ns and peak power up to tens of mW. Although the load resistance $R$ should be larger to increase the conversion efficiency of EOM, the charge of capacitance remains during the operation. For the high-speed operation of time-series information processing such as RNN, the charge of capacitance should be completely discharged until the dataset processed next is input to the system. By adopting an electrical switch to discharge the residual charge, the system can deal with the high-speed input dataset. Furthermore, by using a carrier-injection-based modulator as an EOM, how the signal of EOM remains in the time domain is expected to be controlled by the carrier lifetime and the residual signal can be eliminated by applying the reverse-bias voltage in a carrier-depletion mode.

Next, we discuss the finite pulse response of OEO converter. In the case of finite pulse response, the rise time of charging of EOM by input signal must be taken into account. The voltage of EOM with the rise time is expressed as $V_{\rm{EOM}} = RI_{x}(1-e^{-t_{p}/\tau_{RC}})$. Note that the peak power of input signal $I_{x}$ is enough small with the large road resistance $R$ compared to an impulse response. Totally, the OEO model with the rise time and fall time of pulse response can be written as
\begin{equation}\label{OEO_model_rise_fall}
z(t) = (1-\gamma) \beta x(t) + \alpha z(t-1),
\end{equation}
where $\gamma \equiv \exp(-t_{p}/\tau_{RC})$ is the rise-time RC delayed feedback which $\tau_{RC}\rightarrow0$, $\gamma\rightarrow0$ and $\tau_{RC}\rightarrow\infty$, $\gamma\rightarrow1$. From Eq. \ref{OEO_model_rise_fall}, as the rise time $t_{p}$ is comparably large to the RC delayed feedback $\tau_{RC}$, the charge to EOM is slow and it acts as a loss. Therefore, in the case of the use of finite pulse laser, a loss due to the rise time of charging should be compensated by increasing the loop gain $\beta$. The OEO gain strongly depends on the performance of EOM such as the conversion efficiency, optical loss, and bandwidth. In \cite{nozaki2019}, the maximum conversion efficiency of nanophotonic OEO converter is $2.3\pm0.3$ for a normally-on operation. Although the high-intensity bias light is desirable to further increase the OEO gain, it can induce the additional losses and degrade the efficiency of the EOM device. For example, for the optical modulators with p-n/p-i-n junctions, the reverse-current caused by the absorption of bias light cancels the phase shift to degrade the conversion efficiency of EOM. To avoid this effect, we can use the modulator driven by a voltage via Pockels or Kerr electro-optic effects \cite{Sinatkas2021}. Also unlike p-n/p-i-n junctions, the hybrid MOS optical modulator has the small absorption of bias light and quite low carrier-induced loss \cite{Han2017}. Thus, these high-efficiency OEO converter with the low additional losses can compensate for the decrease of effective OEO gain due to the rise time of charging. On the other hand, the degraded RNN performance due to the loss can be recovered by the accumulation of time-series information due to the fall time of the pulse signal as discussed in Section \ref{analysis}.

\begin{table*}
\caption{\label{OEOtable}The comparison of device parameters of OEO converters.}
\centering
{\renewcommand\arraystretch{1}
\scalebox{1.15}{
\begin{tabular}{c|c|c|c|c|c|c|c}
\hline
 & $V_{\pi}$  & $C$ & $R$ & $\tau_{RC}$ & $t_{p}$ & $I_{x}$ & Bandwidth\\
\hline
\hline
MZM & \begin{tabular}{c} 5 V \end{tabular} & \begin{tabular}{c} 1 pF \end{tabular} & \begin{tabular}{c} 5 k$\Omega$ \end{tabular} & \begin{tabular}{c} 5 ns \end{tabular} & \begin{tabular}{c} 100 ps \end{tabular} & \begin{tabular}{c} 50 mW \end{tabular} & \begin{tabular}{c} $\sim$100 MHz \end{tabular}\\
\hline
Slow-light MZM \cite{Baba2014} & \begin{tabular}{c} 3-5 V \end{tabular} & \begin{tabular}{c} 100 fF \end{tabular} & \begin{tabular}{c} 5 k$\Omega$ \end{tabular} & \begin{tabular}{c} 500 ps \end{tabular} & \begin{tabular}{c} 10 ps \end{tabular} & \begin{tabular}{c} 30-50 mW \end{tabular} & \begin{tabular}{c} $\sim$GHz \end{tabular}\\
\hline
MRR-EOM \cite{Yuan2022} & \begin{tabular}{c} 0.5 V \end{tabular} & \begin{tabular}{c} 5 fF+17 fF \end{tabular} & \begin{tabular}{c} 5 k$\Omega$ \end{tabular} & \begin{tabular}{c} 120 ps \end{tabular} & \begin{tabular}{c} 2 ps \end{tabular} & \begin{tabular}{c} 5 mW \end{tabular} & \begin{tabular}{c} $\sim$10 GHz \end{tabular}\\
\hline
PhC cavity-EOM \cite{nozaki2019} & \begin{tabular}{c} 0.5 V \end{tabular} & \begin{tabular}{c} 5 fF+1 fF \end{tabular} & \begin{tabular}{c} 5 k$\Omega$ \end{tabular} & \begin{tabular}{c} 30 ps \end{tabular} & \begin{tabular}{c} 500 fs \end{tabular} & \begin{tabular}{c} 5 mW \end{tabular} & \begin{tabular}{c} $>$10 GHz \end{tabular}\\

\hline
\end{tabular}
}}
\end{table*}

\section{Conclusion} \label{conclusion}
In summary, this paper addresses optoelectronic RNN using OEO converters with RC delay. We investigated the effect of RC delay on the RNN performance of OE-RNN. It reveals that the RC delay of OEO converters does not degrade the RNN performance rather than can recover the degraded RNN performance due to the loop losses. In theory, we treated the gradient explosion and vanishing problem to analyze the effect of RC delay of OEO converts on the RNN performance. To confirm the analysis, we performed the simulation of time-series information classification task such as MNIST and Fashion-MNIST. The simulated results are in good agreement with the theoretical prediction. We confirm the compensation effect of RC delay for the degraded RNN performance by the gradient explosion or vanishing problem in the large-scale OE-RNN circuit up to $32\times32$. Our proposed OEO-RNN paves a way to a novel optical time-series information processor for a lot of applications such as high-speed optical computing and optical communication with the low latency. 

\section*{Acknowledgment}
We would like to thank S. Sunada at Faculty of Mechanical Engineering, Instite of Science and Engineering, Kanazawa University for the fruitful discussion. We would like to acknowledge T. Uemura for his helpful technical support.

\ifCLASSOPTIONcaptionsoff
  \newpage
\fi

\bibliographystyle{IEEEtran}
\bibliography{IEEEabrv,Bibliography}

\vfill

\end{document}